# High-Receiver-Bandwidth MRI Near Metal Implants with Motion-Consistent Rep2Rep Denoising

Iman Khodarahmi[1], Renqing Luo[1,2,3], Nikola Janjušević[1,2], Christopher J. Burke[1], Shana G. Neumann[1], Yao Wang[3], Li Feng[1,2]

1 Bernard and Irene Schwartz Center for Biomedical Imaging, Department of Radiology, New York University Grossman School of Medicine, New York, NY, USA.
2 Center for Advanced Imaging Innovation and Research (CAI[2]R), Department of Radiology, New York University Grossman School of Medicine, New York, NY, USA.
3 Department of Electrical and Computer Engineering, NYU Tandon School of Engineering, New York, NY, USA

**Word Count:** 5382
**Running Header:** Rep2Rep Denoising for High-Receiver-Bandwidth MRI
**Grant Support:** NIH P41EB017183, R01AR087005

Address correspondence to:
Li Feng, PhD
Center for Advanced Imaging Innovation and Research (CAI2R)
New York University Grossman School of Medicine
660 1st Ave
New York, NY, USA 10016
Email: Li.Feng@nyulangone.org

## Abstract

**Purpose:** To develop and evaluate a self-supervised denoising approach using motion-consistent Rep2Rep learning to enable high-receiver-bandwidth MRI for reduced metal-induced distortion.

**Methods:** Motion-consistent Rep2Rep generates anatomically consistent training repetitions in the presence of inter-repetition motion by combining two acquired repetitions and splitting the combined k-space along the readout direction. The model was trained on 134 clinical low-receiver-bandwidth (300 Hz/pixel) MRI datasets (60 sagittal and 74 axial) and applied to 59 high-receiver-bandwidth (1028 Hz/pixel) MRI datasets (30 sagittal and 29 axial). Three musculoskeletal radiologists independently scored low- and high-receiver-bandwidth images (before and after denoising) in a blinded manner for metal artifact reduction, image sharpness, perceived SNR, and overall image quality (5-1: best to non-diagnostic). A separate blinded reader preference evaluation was performed to compare motion-consistent Rep2Rep with standard Rep2Rep.

**Results:** High-receiver-bandwidth MRI significantly reduced metal-induced artifact, while motion-consistent Rep2Rep improved perceived SNR and overall image quality. Compared to clinical low-receiver-bandwidth MRI, denoised high-receiver-bandwidth MRI achieved significantly higher metal artifact reduction scores (sagittal: 3.69 vs 3.01; axial: 3.83 vs 3.05; both P<0.0001) and comparable overall image quality scores (sagittal: 4.03 vs 3.98, P = 0.5391; axial: 3.95 vs 3.87, P=0.1266). The high-receiver-bandwidth axial protocol also reduced acquisition time by 47%. Motion-consistent Rep2Rep better preserved fine anatomical details than standard Rep2Rep and was preferred in 44 of 59 cases (74.6%).

**Conclusion:** High-receiver-bandwidth MRI with motion-consistent self-supervised Rep2Rep denoising reduces metal-induced artifact without requiring dedicated high-receiver-bandwidth training data. This offers a practical approach for MRI near metal implants, such as postoperative cervical spine imaging.

## 1. Introduction

MRI plays an important role in postoperative evaluation of patients with orthopedic metal implants (1–3). Compared to computed tomography (CT), MRI provides superior soft tissue contrast and enables detailed assessment of anatomical structures and postoperative complications, particularly within or near the spinal canal (4,5). Nevertheless, metal implants continue to pose a significant challenge in MRI because they cause local magnetic field perturbations that result in signal loss, signal pileup, and geometric distortion (3,6). These artifacts can obscure anatomical structures near metal implants and impair diagnosis (3,6,7).

In conventional 2D Cartesian MRI, metal-induced off-resonance primarily disrupts spatial encoding during slice selection and readout, resulting in through-plane and in-plane artifacts such as geometric distortion and signal loss/pileup. Various imaging techniques have been proposed to mitigate metal-induced artifacts (3,6). To reduce in-plane artifacts, increased receiver bandwidth (readout bandwidth) and view-angle tilting (VAT) (8,9) can both reduce off-resonance-induced signal displacement along the frequency-encoding direction(10). To reduce through-plane artifacts, high-bandwidth radiofrequency (RF) pulses are commonly used in many clinical metal artifact reduction protocols (3,6). More advanced techniques, such as slice encoding for metal artifact correction (SEMAC) and multi-acquisition variable-resonance image combination (MAVRIC), provide more comprehensive artifact reduction through additional spatial encoding or spectral acquisitions (11–13). These approaches, however, come with their own limitations. Increasing receiver bandwidth significantly reduces signal-to-noise ratio (SNR) (3,6,10), VAT causes image blurring (9), high-bandwidth RF pulses can increase RF power deposition and specific absorption rate (SAR) (3), and advanced techniques such as SEMAC and MAVRIC require complex acquisitions and longer scan times (1,3,6,14) .

Among these established approaches, increasing receiver bandwidth is particularly attractive because it is simple and can be readily implemented in many MRI sequences without specialized acquisitions. However, although this approach has been recommended (5,15), the associated SNR penalty can negatively affect image quality and diagnostic confidence (15). The SNR loss can be compensated through additional signal averages, but this leads to longer scan times and increased sensitivity to patient motion.

Deep learning-based denoising offers a promising approach to address this limitation and could potentially improve the clinical utility of high-receiver-bandwidth MRI near metal implants (3,16). However, conventional supervised deep learning generally relies on high-SNR reference images for network training, which are difficult to acquire at high receiver bandwidth within clinically reasonable acquisition times. Self-supervised denoising approaches, which do not require high-SNR reference images and have been increasingly applied to various low-SNR MRI applications (17–22) , therefore represent an attractive alternative.

The purpose of this study was to develop and evaluate a self-supervised denoising approach to enable high-receiver-bandwidth (~1000 Hz/pixel, as recommended by Kim et al (15)) MRI with reduced metal-induced artifacts while preserving overall image quality. Our approach is based on the recently developed Repetition-to-Repetition (Rep2Rep) framework (23), a noise-adaptive self-supervised denoising approach based on the Noise2Noise principle (24). Rep2Rep learns from paired MRI repetitions (or signal averages) with shared underlying anatomical structure but independent zero-mean noise realizations without the need for matched high-SNR reference images. In this study, we further developed Rep2Rep for denoising high-receiver-bandwidth MRI near metal implants. We demonstrated that a Rep2Rep model trained using routine clinical low-receiver-bandwidth datasets can be directly applied to high-receiver-bandwidth images with lower SNR to achieve metal artifact reduction. This avoids the need to acquire dedicated high-receiver-bandwidth datasets for network training. More importantly, we developed a motion-consistent Rep2Rep training strategy to address anatomical mismatch between paired repetitions, which can compromise the performance of standard Rep2Rep denoising. The proposed strategy generates more anatomically consistent training repetitions to improve the preservation of fine anatomical details in the denoised images.

The proposed approach was evaluated in postoperative cervical spine MRI of patients with metal instrumentation, where both high spatial resolution and effective metal artifact reduction are important. Our hypothesis was that high-receiver-bandwidth MRI (~1000 Hz/pixel), combined with motion-consistent Rep2Rep denoising trained entirely on routine clinical low-receiver-bandwidth (300 Hz/pixel) data, can reduce metal-induced artifacts while preserving overall image quality.

## 2. Methods

In this section, we first review the original Rep2Rep framework for self-supervised denoising and discuss its limitations associated with inter-repetition motion. The proposed motion-consistent Rep2Rep extension is then presented, followed by the clinical datasets and acquisition protocols, network implementation, and experimental evaluation methods used in this study.

### 2.1 Overview of Rep2Rep Denoising

Rep2Rep is a noise-adaptive self-supervised MRI denoising framework originally developed for low-SNR applications (23) based on the Noise2Noise principle (24). It uses paired MRI repetitions (signal averages) for network training without the need for matched high-SNR reference images, under the assumption that repeated MRI acquisitions share the same underlying anatomical structure but have independent zero-mean noise realizations. Specifically, one noisy repetition serves as the network input, while another repetition serves as the training target. Because random noise in the target repetition is independent of the input repetition, the network can effectively learn to recover the shared anatomical structure but not the noise through minimization of a mean squared error (MSE) loss between the network output and target image. Rep2Rep uses a convolutional dictionary learning network (CDLNet) with an explicit noise-adaptive mechanism that incorporates noise information to accommodate different noise levels between training and inference (25,26). This allows the trained model to perform denoising across different SNR levels without the need for retraining or fine-tuning.

Rep2Rep is also compatible with parallel MRI techniques such as GRAPPA (GeneRalized Autocalibrating Partially Parallel Acquisitions) (27), which enables denoising of images reconstructed from accelerated MRI acquisitions (23). Consider two repetitions acquired with complementary interleaved undersampling patterns, each with 2-fold acceleration (R = 2), as in our clinical sagittal MRI protocol described below. In standard Rep2Rep training, the two repetitions are reconstructed separately using GRAPPA, followed by fast Fourier transform (FFT) and sensitivity-weighted coil combination to generate two complex-valued images, which are used as the network input and target (**Supporting Information Figure S1**). The complex-valued representation is needed here to preserve the zero-mean noise characteristics required for Noise2Noise learning (23,24).

In our previous work (23), Rep2Rep demonstrated superior denoising performance compared with Monte Carlo Stein's unbiased risk estimator (MC-SURE) a self-supervised denoising approach that does not require paired acquisitions (20). However, an important limitation of Rep2Rep is its dependence on anatomical consistency between repeated MRI acquisitions. In many applications, inter-repetition motion is difficult to avoid, which can introduce anatomical mismatch between the network input and target and therefore violate the shared-signal assumption of Noise2Noise. In this case, the network may treat these structural differences as noise and suppress fine anatomical details, thus leading to image blurring. This limitation is particularly relevant to motion-sensitive applications such as cervical spine MRI, where involuntary neck movement and swallowing may cause anatomical difference between repeated acquisitions.

### 2.2 Motion-Consistent Rep2Rep Training

To address the limitation of standard Rep2Rep due to inter-repetition motion, we propose a motion-consistent extension of Rep2Rep based on k-space splitting along the readout direction ($k_x$), referred to as motion-consistent Rep2Rep, as illustrated in **Figure 1a**. In this approach, readout oversampling is first removed from each repetition. The resulting multicoil k-space data from the two repetitions are then combined to form a fully sampled k-space dataset. The two repetitions can be acquired either with complementary interleaved undersampling patterns (as in our clinical sagittal MRI protocol) or with full k-space sampling in each repetition (as in our clinical axial MRI protocol). The combined data are then uniformly divided into two complementary subsets along the readout direction, with one subset containing odd-indexed samples and the other subset containing even-indexed samples. In this way, each subset corresponds to 2-fold uniform undersampling along the readout direction. GRAPPA reconstruction is then applied separately to each subset to recover the missing k-space data, followed by FFT and sensitivity-weighted coil combination to generate paired complex-valued images. The two reconstructed images are subsequently used as the network input and target for self-supervised Rep2Rep training.

The rationale for the readout-direction split is that all samples along each readout (frequency-encoding) line are acquired during the same readout acquisition, which occurs within a much shorter temporal window than the acquisition of different phase-encoding lines or separate repetitions. In addition, removal of readout oversampling helps reduce potential noise correlations between adjacent readout samples caused by vendor-implemented anti-aliasing filters (28) (see **Supporting Information Figure S2**). Therefore, the odd- and even-indexed readout samples are expected to share the same anatomical information, even in the presence of inter-repetition motion. Because both subsets are generated from the same combined k-space dataset, they also share similar motion-related inconsistencies across phase-encoding lines or signal repetitions. As a result, the two reconstructed images are expected to have reduced anatomical mismatch compared to images reconstructed from separate repetitions. It should be noted that the proposed strategy aims to generate motion-consistent image repetitions for Rep2Rep training rather than explicitly correct for motion artifacts. In the presence of inter-repetition motion, motion-induced blurring caused by the combination of repetitions can still propagate into the reconstructed images. However, this issue also exists in conventional clinical MRI protocols that combine multiple signal averages.

### 2.3 MRI Datasets and Imaging Protocols

The study was approved by the Institutional Review Board (IRB). All MRI datasets were acquired on a 1.5T clinical MRI scanner (MAGNETOM Aera, Siemens Healthineers, Erlangen, Germany) using T2-weighted 2D fast spin-echo (FSE) sequences in patients with cervical spinal instrumentation (interbody spacers, anterior plate-screw constructs, and posterior screw-rod constructs).

MRI datasets used for network training were retrospectively collected from routine clinical scans. The requirement for individual informed consent was waived due to the retrospective nature of this part of the study. A total of 60 sagittal and 74 axial datasets from 74 patients (46 male, 28 female, age = 59 ± 15 years) were collected for network training. Raw k-space data were automatically collected using the Yarra toolbox developed at our institution (29). 14 sagittal datasets were unavailable due to data loss during the automated collection.

The clinical protocol included both T2-weighted sagittal and axial sequences. Sequence parameters for sagittal imaging included: FOV = 220 × 220 $mm^2$, matrix size = 256 × 256, in-plane spatial resolution = 0.9 × 0.9 $mm^2$, phase oversampling = 85%, slice thickness = 3 mm, 17 slices, repetition time/echo time (TR/TE) = 3000/96 ms, echo spacing = 7.96 ms, turbo factor = 13, refocusing flip angle = 170°, two complementary interleaved signal averages, each with 2-fold undersampling (R = 2, see Figure 1), phase-encoding direction = head-to-foot, receiver bandwidth = 300 Hz/pixel, and total scan time = 1 min 59 s. Instead of acquiring a single fully sampled repetition, our clinical sagittal protocol acquired two complementary R = 2 repetitions, which were reconstructed separately and then averaged in the magnitude image domain to reduce sensitivity to patient motion.

Sequence parameters for axial imaging included: FOV = 180 × 160 $mm^2$, matrix size = 256 × 230, in-plane spatial resolution = 0.7 × 0.7 $mm^2$, phase oversampling = 65%, slice thickness = 3 mm, 30 slices, TR/TE = 4000/91 ms, echo spacing = 9.12 ms, turbo factor = 11, refocusing flip angle = 150°, two fully sampled signal averages, phase-encoding direction = anterior-to-posterior, receiver bandwidth = 300 Hz/pixel, and total scan time = 3 min 58 s. Our clinical axial protocol acquired two fully sampled signal averages to ensure sufficient SNR at the higher in-plane spatial resolution.

The independent testing cohort comprised 30 patients (22 male, 8 female, age = 66 ± 14.5 years) who underwent additional prospective high-receiver-bandwidth MRI at the end of their routine clinical exams after providing informed consent. A total of 30 sagittal and 29 axial high-receiver-bandwidth datasets were available for evaluation. One axial raw data was lost during automated Yarra data transfer. The high-receiver-bandwidth imaging parameters were similar to those of the corresponding clinical protocols, except for the receiver bandwidth and associated changes in TE and echo spacing, as well as acceleration for axial imaging. Specifically, the receiver bandwidth was increased from 300 to 1028 Hz/pixel for both sagittal and axial acquisitions. For sagittal imaging, the TE and echo spacing were reduced to 81 ms and 7.32 ms, respectively, with the same scan time of 1 min 59 s. For axial imaging, the TE and echo spacing were reduced to 81 ms and 7.40 ms, respectively. In addition, two complementary interleaved repetitions with 2-fold undersampling were acquired for the high-receiver-bandwidth axial protocol, which reduced the scan time from 3 min 58 s to 2 min 06 s compared to the routine clinical axial protocol. Matched clinical datasets acquired at lower receiver bandwidth (300 Hz/pixel) in this evaluation cohort were also collected, and they were only used for image quality comparisons and were not included in network training.

**2.4 Network Training and Inference**

All denoising models were implemented using the Rep2Rep framework described in Section 2.1, with details of the network architecture provided in our previous work (23,25,30). The 60 sagittal and 74 axial training datasets acquired at 300 Hz/pixel were pooled together to train a single joint model that could be applied to both sagittal and axial MRI, despite differences in spatial resolution and acquisition parameters. The joint model was used mainly to simplify implementation and avoid the need for separate orientation-specific models. The network was implemented using complex-valued 3D CDLNet (30) with 30 unrolled iterations. Model training used a batch size of 1 and the Noise2Noise MSE loss. The Adam optimizer (31) was used for network training with a learning rate of $5 \times 10^{-5}$. The model was trained for 500 epochs on a single NVIDIA A100 GPU, with a total training time of ~35 hours.

For inference, the trained model was directly applied to the high-readout-bandwidth datasets (1028 Hz/pixel) without k-space splitting. The complete k-space dataset combined from the two repetitions was simply reconstructed using FFT and sensitivity-weighted coil combination to generate complex-valued images, as shown in **Figure 1b**.

For both training and inference, coil sensitivity maps were estimated from low-resolution multi-coil images reconstructed from the central k-space region using the Walsh method (32). Spatially varying noise maps used for noise-adaptive training and inference were estimated directly from the multicoil images as described previously (23) and incorporated into the model to accommodate differences in SNR between training and inference. Because MSE-based denoising may lead to image smoothness and loss of fine structural details (33), the denoised images were additionally processed with a sharpening step and a dithering procedure to improve image sharpness and restore a more natural noise texture, as implemented in our original Rep2Rep study (23). Specifically, sharpening was performed by adding a scaled high-frequency component to the denoised image, while dithering was performed by adding a fraction of the residual between the original noisy input and denoised image to the sharpened image, as shown in **Figure 2**.

### 2.5 Experimental Design and Image Quality Assessment

Different experiments were performed to assess the performance of the proposed approach, including the benefit of motion-consistent Rep2Rep training, the image quality of high-receiver-bandwidth MRI, and the effect of noise adaptation on denoising performance.

First, to evaluate the benefit of the proposed motion-consistent Rep2Rep training, a blinded pairwise preference study was performed to compare motion-consistent Rep2Rep (**Figure 1**) with standard Rep2Rep (**Supporting Information Figure S1**). For each high-receiver-bandwidth dataset, the two denoised images were anonymized and presented side by side in a randomized order to an experienced musculoskeletal radiologist (I.K.), who selected the preferred image based on overall image quality and preservation of fine anatomical details. Both methods were applied to the same high-receiver-bandwidth data with the same post-processing procedures. The results were reported as the number and percentage of cases in which each method was preferred.

Second, to evaluate the image quality of high-receiver-bandwidth MRI after motion-consistent Rep2Rep denoising, three matched image sets were compared for each patient and imaging orientation: (1) clinical low-receiver-bandwidth images, (2) high-receiver-bandwidth images without denoising, and (3) high-receiver-bandwidth images after motion-consistent, noise-adaptive Rep2Rep denoising. All images were anonymized and pooled for independent evaluation by three experienced musculoskeletal radiologists (I.K., C.J.B., and S.G.N.), who were blinded to the imaging information. Each image was scored based on four metrics (metal artifact reduction, image sharpness, perceived SNR, and overall image quality) using a five-point Likert scale (5 = excellent, 4 = good, 3 = adequate, 2 = poor, and 1 = non-diagnostic). Reader scores were averaged and reported as mean ± standard deviation for each metric. Pairwise differences between images were assessed using a two-sided Wilcoxon signed-rank test ($P<0.05$), and inter-reader agreement was assessed using mean pairwise quadratic-weighted Cohen's kappa coefficients.

Third, to investigate the effect of noise adaptation given the substantial difference in SNR levels between training and inference, we compared two Rep2Rep models, one with noise adaptation and one without. Both models were trained using the same clinical low-bandwidth datasets and the proposed motion-consistent training repetitions, with the same network architecture and training parameters except for noise adaptation. The trained models were applied to the same high-bandwidth datasets, and their image quality was compared both before and after sharpening and dithering to assess whether noise adaptation provides additional benefits in the final images. As an additional comparison, joint training using both sagittal and axial datasets was evaluated against separate orientation-specific training to assess whether a single joint model could accommodate differences in imaging orientations and acquisition parameters without affecting image quality.

## 3. Results

### 3.1 High Receiver-Bandwidth Cervical Spine MRI with Reduced Metal Artifacts

Representative cervical spine images comparing clinical low-receiver-bandwidth (300 Hz/pixel) with high-receiver-bandwidth (1028 Hz/pixel) imaging without and with denoising using the proposed motion-consistent Rep2Rep learning are shown in **Figures 3-5**. For sagittal imaging (**Figure 3**), increasing the receiver bandwidth reduced metal-induced geometric distortion and signal pileup near the implants (red arrows). However, the high-bandwidth images before denoising suffered from substantially lower SNR and degraded image quality. Motion-consistent Rep2Rep effectively improved perceived SNR while preserving the reduced metal-induced artifacts with improved visualization of anatomical structures near the implants.

Similar results were observed on axial imaging of different patients (**Figures 4 and 5**), where the increased receiver bandwidth reduced metal-induced artifacts but resulted in substantially lower SNR, particularly given the higher spatial resolution acquired in the axial protocol. Rep2Rep denoising substantially improved image quality and anatomical visualization. Importantly, compared with the clinical low-receiver-bandwidth axial protocol that did not use parallel imaging, our high-receiver-bandwidth axial protocol enabled 2-fold GRAPPA acceleration to reduce the acquisition time from 3 min 58 s down to 2 min 06 s.

### 3.2 Evaluation of Motion-Consistent Rep2Rep Against Standard Rep2Rep

**Figure 6** compares the anatomical consistency of training repetitions generated using the original two signal averages and the proposed readout-direction k-space split. In both sagittal and axial MRI, the original repetitions showed clear anatomical mismatch, as shown in the difference maps. In contrast, the proposed motion-consistent repetitions showed substantially less structural differences. These results suggest that the proposed readout-direction split can effectively reduce anatomical mismatch between the training repetitions, therefore better satisfying the shared-signal assumption of the Noise2Noise principle.

**Figure 7** shows representative comparisons between standard Rep2Rep trained using the original two repetitions and the proposed motion-consistent Rep2Rep. Although both approaches effectively reduced image noise, standard Rep2Rep showed image blurring and loss of fine anatomical details in different regions, as indicated by the yellow arrows. In contrast, motion-consistent Rep2Rep better preserved anatomical boundaries and fine structural details in both sagittal and axial MRI. For this comparison, both standard Rep2Rep and motion-consistent Rep2Rep images were processed using the same sharpening and dithering procedures to ensure a fair comparison. These results suggest that improved anatomical consistency between training repetitions can reduce image blurring associated with inter-repetition motion in Rep2Rep denoising.

### 3.3 Radiologist Evaluation of Image Quality

Motion-consistent Rep2Rep was preferred in 17 of 30 sagittal cases (56.7%) and 27 of 29 axial cases (93.1%), totaling 44 of 59 cases (74.6%) overall. The two methods were rated as equivalent in the remaining 15 cases (25.4%), including 2 axial and 13 sagittal cases. Standard Rep2Rep was not preferred over motion-consistent Rep2Rep in any case.

The results of the reader evaluation are summarized in **Figure 8**. For sagittal imaging, the metal artifact reduction scores (mean ± SD) were 3.01 ± 0.35 for clinical low-receiver-bandwidth images, 3.80 ± 0.40 for high-receiver-bandwidth images without denoising ($P < 0.0001$ vs clinical MRI), and 3.69 ± 0.47 for high-receiver-bandwidth images after Rep2Rep denoising ($P < 0.0001$ vs clinical MRI and $P = 0.0016$ vs images without denoising). For axial imaging, the corresponding scores were 3.05 ± 0.30, 3.87 ± 0.33, and 3.83 ± 0.38, respectively ($P < 0.0001$ for both high-receiver-bandwidth images vs clinical low-receiver-bandwidth images and $P = 0.2482$ for denoised vs non-denoised high-receiver-bandwidth images). These results demonstrated the effectiveness of increased receiver bandwidth in reducing metal-induced artifacts.

For image sharpness, sagittal imaging received scores of 3.99 ± 0.11 for clinical low-receiver-bandwidth images, 3.76 ± 0.43 for high-receiver-bandwidth images without denoising ($P < 0.0001$ vs clinical MRI), and 4.02 ± 0.30 for high-receiver-bandwidth images after Rep2Rep denoising ($P < 0.0001$ vs images without denoising; $P = 0.6426$ vs clinical MRI). For axial imaging, the corresponding scores were 3.94 ± 0.38, 3.57 ± 0.52, and 3.79 ± 0.44, respectively ($P < 0.0001$ for clinical vs non-denoised high-receiver-bandwidth images and for denoised vs non-denoised high-receiver-bandwidth images; $P = 0.0202$ for denoised high-receiver-bandwidth images vs clinical MRI).

For perceived SNR, the scores for sagittal imaging were 3.97 ± 0.18 for clinical low-receiver-bandwidth images, 3.21 ± 0.44 for high-receiver-bandwidth images without denoising ($P < 0.0001$ vs clinical MRI), and 4.09 ± 0.29 for images after Rep2Rep denoising ($P < 0.0001$ vs images without denoising and $P = 0.0009$ vs clinical MRI). For axial imaging, the corresponding scores were 3.87 ± 0.33, 2.94 ± 0.32, and 4.10 ± 0.31, respectively ($P < 0.0001$ for all paired comparisons).

For overall image quality, the scores for sagittal imaging were 3.98 ± 0.21 for clinical low-receiver-bandwidth images, 3.66 ± 0.48 for high-receiver-bandwidth images without denoising (P < 0.0001 vs clinical MRI), and 4.03 ± 0.23 for high-receiver-bandwidth images after Rep2Rep denoising (P < 0.0001 vs images without denoising; P = 0.5391 vs clinical MRI). For axial imaging, the corresponding scores were 3.87 ± 0.43, 3.03 ± 0.36, and 3.95 ± 0.30, respectively (P < 0.0001 for non-denoised high-receiver-bandwidth images vs clinical MRI and for denoised vs non-denoised high-receiver-bandwidth images; P = 0.1266 for denoised high-receiver-bandwidth images vs clinical MRI). High-receiver-bandwidth images with Rep2Rep denoising achieved overall image quality scores that were not significantly different from those of clinical low-receiver-bandwidth images, while yielding significantly improved metal artifact reduction.

For inter-reader agreement across all three readers, the mean pairwise quadratic-weighted Cohen's kappa coefficients were 0.515 for metal artifact reduction, 0.128 for image sharpness, 0.655 for perceived SNR, and 0.508 for overall image quality. These results indicated moderate to good agreement for metal artifact reduction, perceived SNR, and overall image quality. The agreement for image sharpness scoring was relatively low, which is likely due to the limited variability of reader scores, which were mostly concentrated around 4 on the five-point scale.

### 3.4 Additional Evaluations of Rep2Rep

**Figure 9** compares noise-adaptive and non-adaptive Rep2Rep models trained on the same clinical low-receiver-bandwidth datasets and applied to high-receiver-bandwidth MRI. Before post-processing, the non-adaptive results showed residual noise compared to the noise-adaptive results. However, after sharpening and dithering, which was optimized separately for them, the two models yielded visually comparable images in both orientations. These results suggest that the additional visual benefit of noise adaptation may be less apparent after post-processing for this particular application.

Given that post-processing reduced the visible differences between noise-adaptive and non-adaptive Rep2Rep, we further investigated whether the reduced anatomical detail observed with standard Rep2Rep could be recovered through stronger image sharpening (**Supporting Information Figure S3**). The results show that although increased sharpening enhanced the apparent sharpness of standard Rep2Rep images, it did not fully restore the fine anatomical structures preserved by motion-consistent

Rep2Rep. This suggests that the benefit of motion-consistent training cannot be reproduced simply by applying stronger sharpening to standard Rep2Rep.

Finally, **Figure 10** compares joint training using both sagittal and axial datasets with separate orientation-specific training. The two training strategies yielded visually comparable denoised images despite the differences in spatial resolution and acquisition parameters between orientations. These results demonstrate the feasibility of using a single joint Rep2Rep model for both orientations, which can simplify network training and potential clinical translation.

## 4. Discussion

This study proposed an approach for MRI near metal implants that combines high-receiver-bandwidth acquisition with self-supervised denoising using a motion-consistent Rep2Rep framework. Our results demonstrated the feasibility of this approach for reducing metal-induced artifacts while improving image quality in cervical spine MRI. In this section, we discuss the clinical value of high-receiver-bandwidth MRI, the advantages of self-supervised denoising and motion-consistent training, and the potential clinical applications and limitations of the current study.

### 4.1 Clinical Value of High-Receiver-Bandwidth MRI

Metal-induced artifacts remain a major challenge in postoperative cervical spine MRI, where small anatomical structures such as the spinal cord and nerve roots require high spatial resolution for accurate delineation. Although advanced metal artifact reduction techniques such as VAT, SEMAC, and MAVRIC can reduce these artifacts and have been proved particularly valuable for patients with large metallic prostheses (1–3,6), they have different limitations, such as acquisition complexity and long scan times, which restrict routine clinical use in postoperative spine MRI.

Increasing receiver bandwidth represents a simple and practical approach to metal artifact reduction because it can be implemented directly in routine clinical MRI protocols without specialized pulse sequences. Recent reviews have emphasized the importance of high-receiver-bandwidth MRI for postoperative spine evaluation (5,15). In particular, Kim et al. recommended a readout bandwidth of ~1000 Hz/pixel for FSE MRI protocols in postoperative spine imaging (15). Higher readout bandwidth reduces off-resonance-induced spatial displacement near metal implants, albeit at the cost of reduced SNR, which can reduce image quality, particularly at high spatial resolution. This SNR tradeoff has long represented a major challenge for the clinical use of high-receiver-bandwidth MRI despite its practical advantages.

Our results demonstrated that the SNR reduction associated with high-receiver-bandwidth MRI can be effectively addressed using self-supervised Rep2Rep denoising. The combination of high receiver bandwidth and denoising achieved significantly reduced metal-induced artifacts while maintaining overall image quality comparable to clinical low-receiver-bandwidth MRI. In addition, our high-receiver-bandwidth axial MRI protocol enabled 2-fold GRAPPA acceleration to halve the acquisition time compared to our routine clinical protocol. These findings suggest that high-receiver-bandwidth MRI combined with self-supervised denoising may provide a practical and effective approach for routine postoperative cervical spine MRI.

### 4.2 Motion-Consistent Rep2Rep Training

The original Rep2Rep framework is based on the Noise2Noise principle to employ two independently acquired repetitions as the input and target for network training, under the assumption that they share the same underlying anatomical information but have independent zero-mean noise. However, any anatomical mismatch between repeated acquisitions due to patient motion, such as involuntary swallowing and neck movement in cervical spine MRI, can violate this assumption. These anatomical differences can be incorrectly treated as noise during network training, thus potentially leading to image blurring and loss of fine anatomical details after Rep2Rep denoising.

In this work, we proposed a motion-consistent training strategy to address this limitation by generating anatomically consistent training repetitions through readout-direction k-space splitting. Instead of independently reconstructing the two acquired repetitions, the proposed approach first combines their k-space data, either from complementary interleaved undersampled acquisitions or fully sampled repetitions, and then uniformly divides the combined data into even and odd samples along the readout direction. In this way, both subsets contain data from the same phase-encoding acquisitions and therefore share the same temporal acquisition effects, including any motion between repetitions. This strategy effectively generates motion-consistent training repetitions with reduced anatomical mismatch, thereby better satisfying the underlying assumption of Noise2Noise learning.

The proposed motion-consistent training strategy achieved reduced anatomical mismatch compared with the original repetition pairs (**Figure 6**), while motion-consistent Rep2Rep better preserved fine anatomical details than standard Rep2Rep in both sagittal and axial imaging (**Figure 7**). Moreover, we showed that increased image sharpening could not fully restore the anatomical details lost with standard Rep2Rep (**Supporting Information Figure S3**). These findings suggest that anatomical consistency plays an important role in preserving fine structural details in Rep2Rep denoising and demonstrate the benefit of the proposed motion-consistent training strategy.

### 4.3 Noise-Adaptive Rep2Rep for High-Bandwidth MRI

An important component of the original Rep2Rep framework is its noise-adaptive capability, which allows the denoising model to accommodate different noise levels between the training and inference datasets. This is particularly relevant to our study, as our Rep2Rep models were trained using routine clinical low-receiver-bandwidth MRI data only and were applied directly to high-receiver-bandwidth MRI with substantially lower SNR. Our results showed that Rep2Rep without noise adaptation has residual noise (**Figure 9**), which is likely due to the higher noise level in the evaluation data with a higher receiver bandwidth compared to the training data.

Interestingly, despite the differences in the direct network outputs, both models yielded visually comparable final images after post-processing that was optimized separately for each model. As discussed previously, one limitation of MSE-based denoising is its tendency to yield certain smoothing in the denoised images with reduced delineation of fine anatomical details (33). Therefore, additional sharpening and dithering were applied to improve apparent image sharpness and restore a more natural noise texture, as we proposed in our original Rep2Rep study (23) and demonstrated in **Figure 2**. Since dithering deliberately reintroduces a certain level of noise into the denoised images, the difference in residual noise between the two models becomes smaller in the final images. This observation suggests that noise adaptation may be less critical for this particular application where additional post-processing is applied. However, this should not be interpreted as evidence that noise adaptation is generally unnecessary. For example, our previous study demonstrated that when the noise level of evaluation data was lower than that of the training data, non-adaptive denoising resulted in increased image blurring instead of residual noise (23).

### 4.4 Clinical Applications and Further Reduction of Scan Time

The proposed approach demonstrated different advantages for sagittal and axial cervical spine MRI based on our clinical protocols. In sagittal imaging, increased receiver bandwidth reduced metal-induced artifacts without increasing acquisition time, and both the clinical and high-receiver-bandwidth protocols used 2-fold parallel imaging. In contrast, the clinical axial protocol used a higher spatial resolution, and parallel imaging was therefore not used for SNR consideration. Rep2Rep denoising allowed us to use 2-fold GRAPPA acceleration in high-receiver-bandwidth axial imaging, reducing the acquisition time from 3 min 58 s down to 2 min 06 s. Therefore, Rep2Rep not only addresses the SNR penalty associated with high-receiver-bandwidth MRI but also provides additional flexibility for further scan-time reduction.

Although evaluated in postoperative cervical spine MRI only, the proposed approach has potential applications in other anatomical regions with metal implants. In addition, high-receiver-bandwidth MRI combined with self-supervised denoising could also be used in combination with other techniques such as VAT, SEMAC, or MAVRIC for further improved metal artifact reduction. Furthermore, the motion-consistent Rep2Rep training strategy is not limited to MRI near metal implants and could potentially be applied to other applications where inter-repetition motion limits the performance of standard Rep2Rep denoising.

### 4.5 Limitations of the Study

This study has several limitations that warrant discussion. First, our Rep2Rep models were trained on routine clinical low-receiver-bandwidth MRI data only. Although our results demonstrated the feasibility of applying these models to high-receiver-bandwidth MRI with substantially different noise levels, a direct comparison with models trained using matched high-receiver-bandwidth data was not performed due to the lack of sufficient high-receiver-bandwidth training data, which would require additional dedicated acquisitions. Second, the proposed approach was evaluated on T2-weighted cervical spine MRI only. Although the motion-consistent training strategy is not specific to T2-weighted imaging, its performance on other image contrasts, anatomical regions, and acquisition protocols requires further investigation. Third, while our reader study demonstrated improvements in metal artifact reduction, perceived SNR, and overall image quality, diagnostic accuracy was not directly assessed, as the present study focused on evaluation of image quality in T2-weighted MRI. Further investigation is warranted to evaluate whether these improvements can be translated into better detection and characterization of postoperative abnormalities.

## 5. Conclusion

This study demonstrated the feasibility of high-receiver-bandwidth MRI near metal implants combined with self-supervised motion-consistent Rep2Rep denoising. The proposed approach effectively reduced metal-induced artifacts while preserving image quality comparable to conventional clinical MRI, with the potential for further reduction of scan time. In addition, the proposed motion-consistent Rep2Rep training strategy improved the preservation of fine anatomical details compared with standard Rep2Rep.

## Code Availability

The source code for Rep2Rep denoising is publicly available at: https://github.com/RapidMRI-NYU/Rep2Rep

## Acknowledgment

This work was supported in part by the NIH (P41EB017183 and R01AR087005) and was performed under the rubric of the Center for Advanced Imaging Innovation and Research (CAI$^2$R), an NIBIB National Center for Biomedical Imaging and Bioengineering.

ChatGPT (OpenAI, CA, USA) was used to assist with grammar correction and improvement of readability.

## Figures

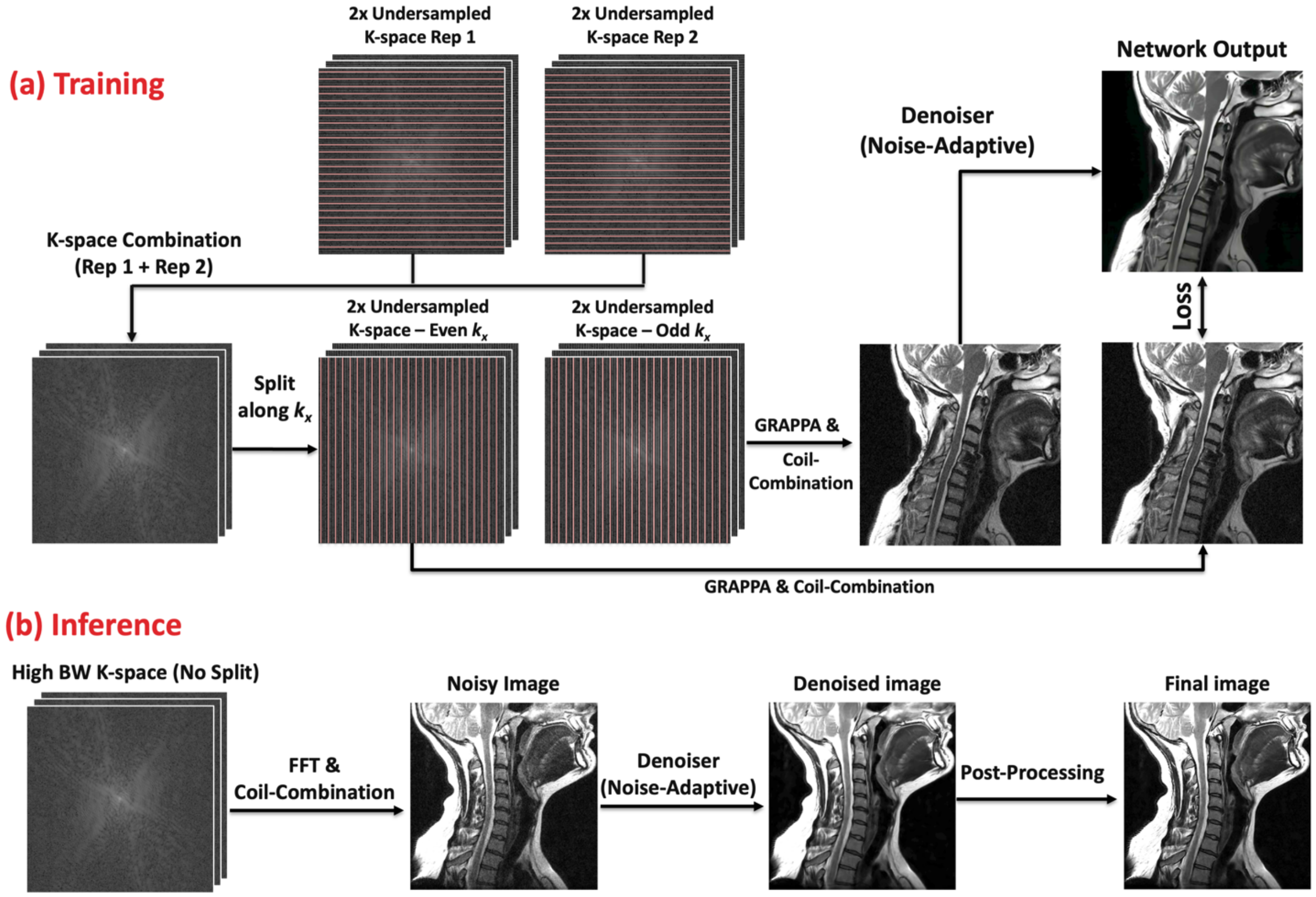

**Figure 1:** (a) The training workflow of the proposed motion-consistent Rep2Rep denoising approach. Two complementary interleaved, undersampled k-space repetitions (R = 2) are first combined to form fully sampled k-space, which is then divided into odd and even samples along the readout direction. The two subsets are independently reconstructed using GRAPPA and coil combination to generate motion-consistent image repetitions for self-supervised training. This workflow can also be applied to fully sampled k-space repetitions, which can be directly combined before the readout-direction k-space split. (b) The inference workflow of motion-consistent Rep2Rep denoising. High-receiver-bandwidth k-space data are reconstructed without k-space splitting and processed using the trained denoising model. The denoised images are then post-processed using sharpening and dithering to generate the final images.

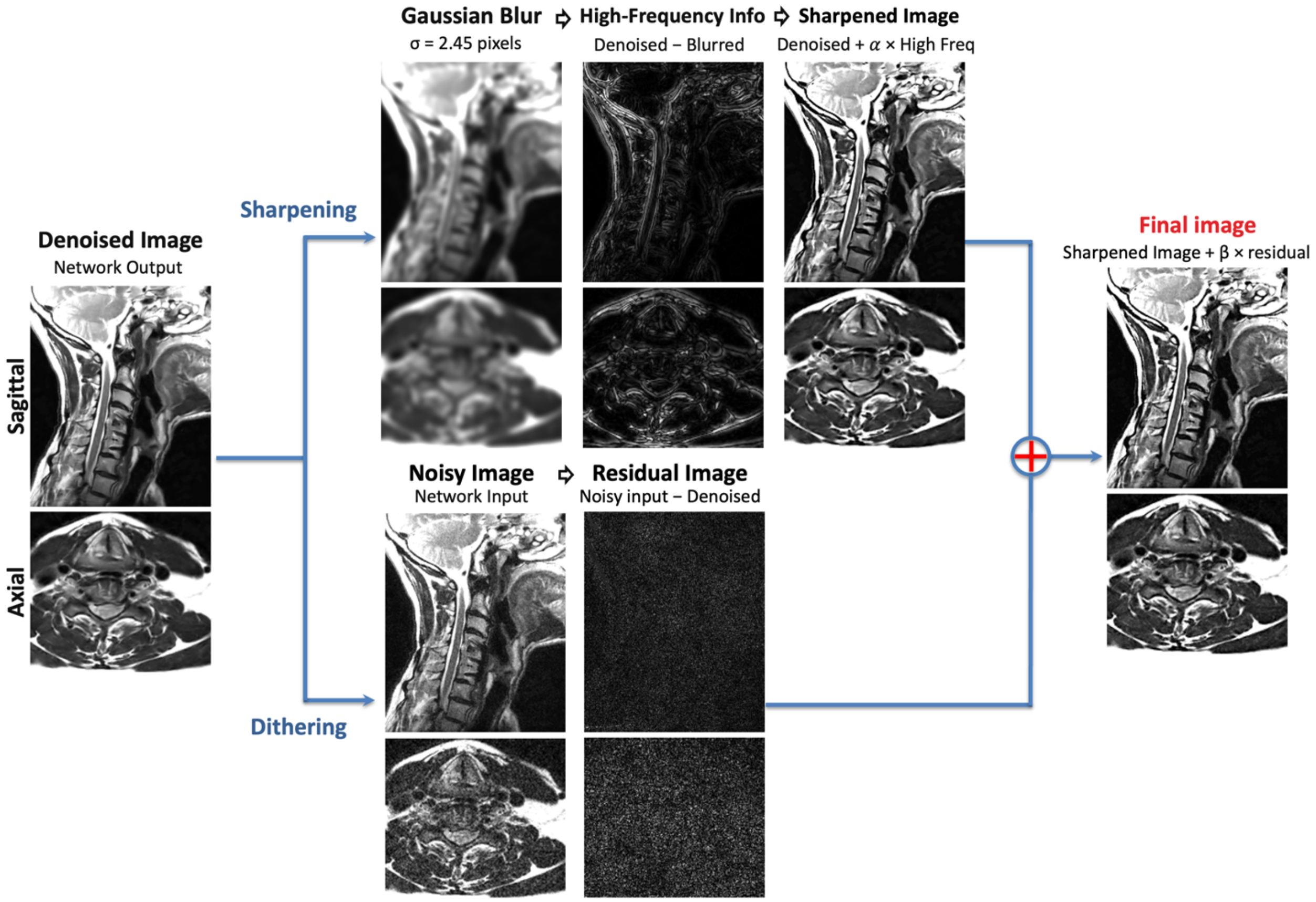

**Figure 2:** Representative sagittal and axial cervical spine MRI examples showing the postprocessing procedures applied to denoised images. Image sharpening was performed by adding a scaled high-frequency component (α = 0.6 for both sagittal and axial), calculated as the difference between the denoised image and its Gaussian-blurred version (Gaussian kernel standard deviation, σ = 2.45 pixels). Dithering was subsequently applied by adding a fraction (β = 0.4 for sagittal and 0.25 for axial) of the residual between the original noisy input and denoised image to restore a more natural noise texture in the final image. The parameters were empirically selected to achieve the best visual appearance and were kept fixed across all datasets within each imaging orientation.

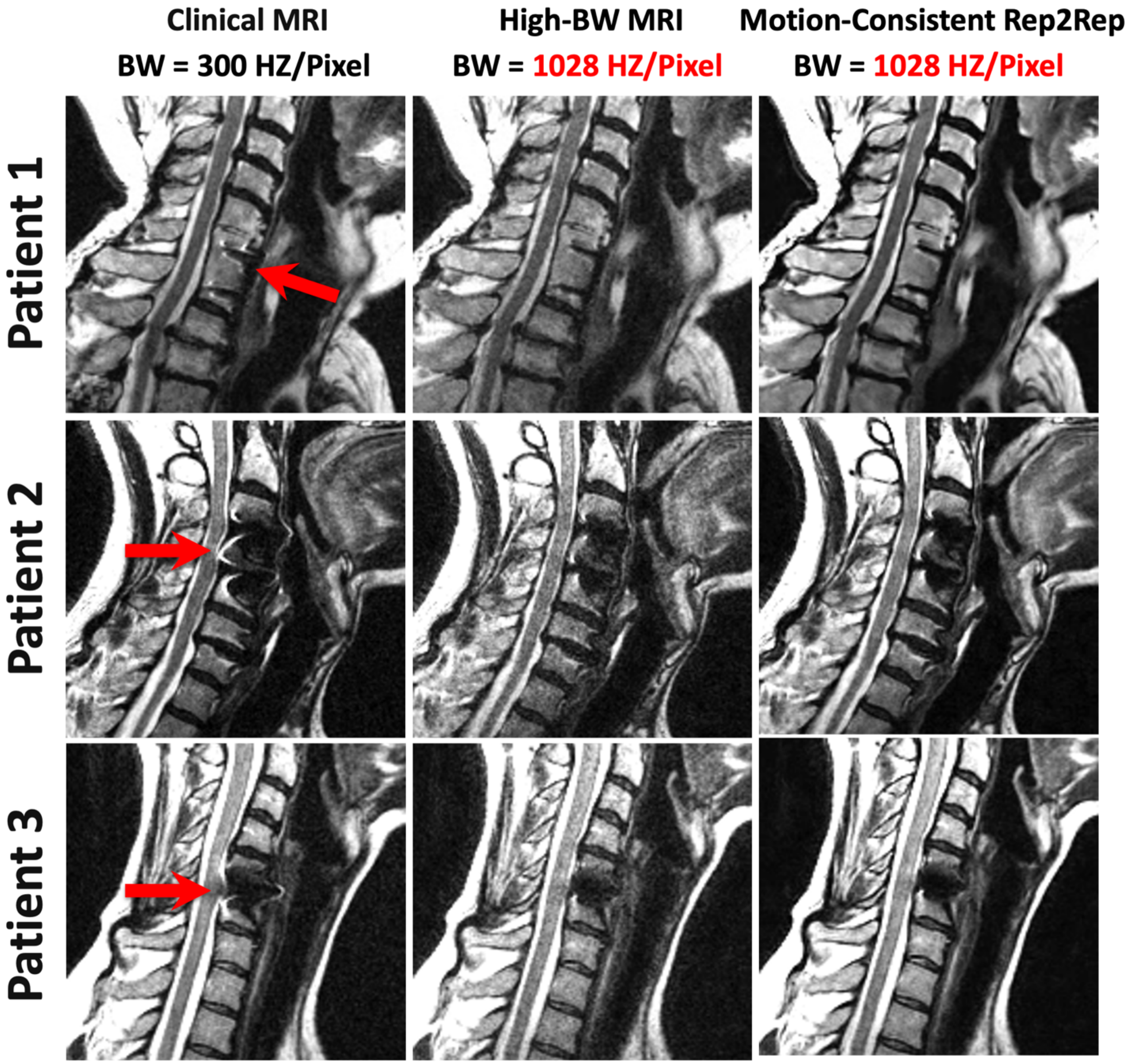

**Figure 3:** Representative sagittal cervical spine MRI images from three patients with metal implants. Clinical MRI acquired at 300 Hz/pixel (left column) is compared with high-receiver-bandwidth MRI acquired at 1028 Hz/pixel before (middle column) and after (right column) motion-consistent Rep2Rep denoising. Increased readout bandwidth reduced metal-induced geometric distortion and signal pileup, as indicated by the red arrows, but resulted in reduced SNR (middle column). Motion-consistent Rep2Rep improved image quality while preserving the reduced metal-induced distortion.

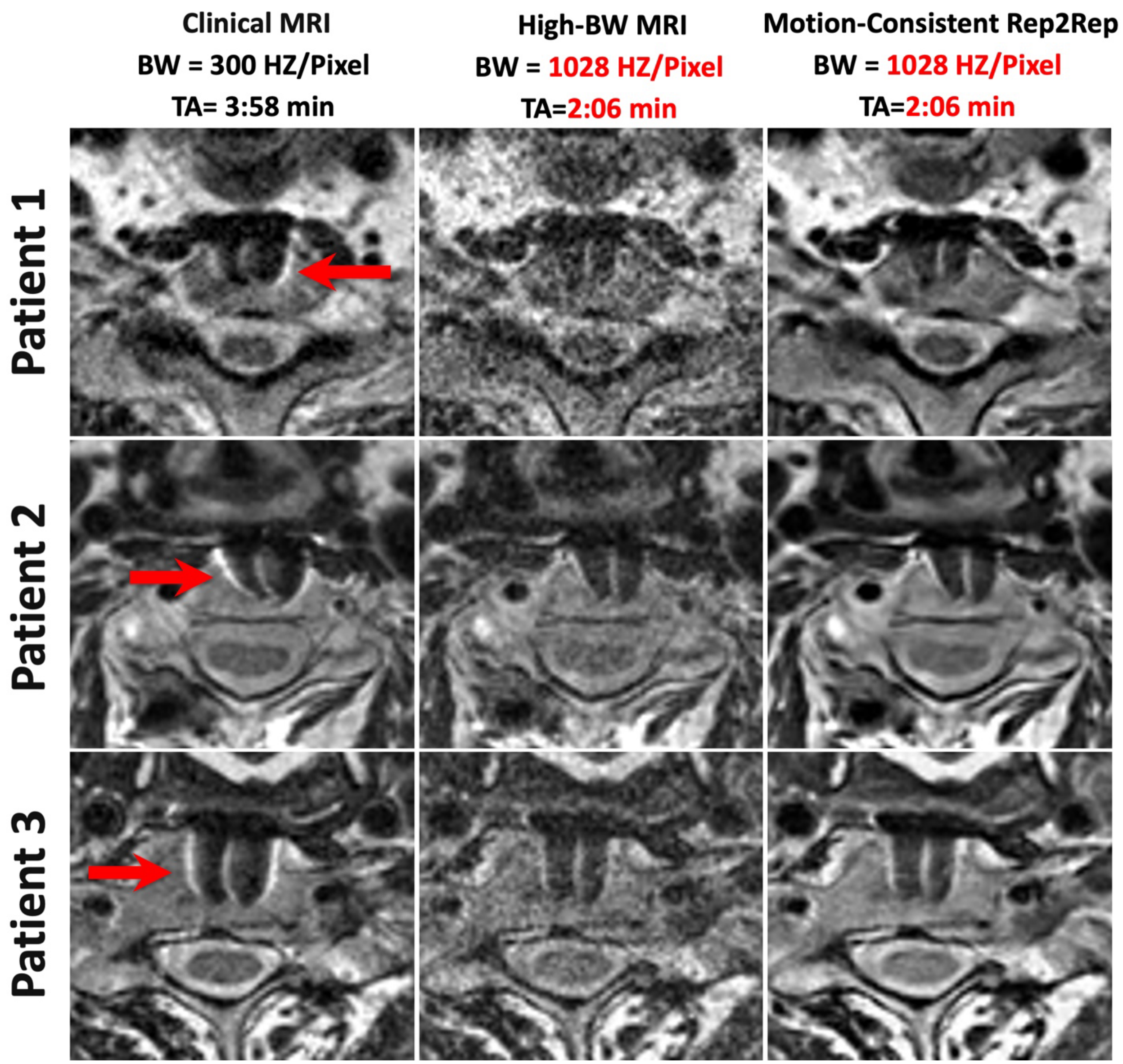

**Figure 4:** Representative axial cervical spine MRI images from three patients with metal implants. Clinical MRI acquired at 300 Hz/pixel (left column) is compared with high-receiver-bandwidth MRI acquired at 1028 Hz/pixel before (middle column) and after (right column) motion-consistent Rep2Rep denoising. The red arrows indicate regions affected by metal-induced artifacts on clinical images. High readout bandwidth reduced these artifacts at the expense of lower SNR, while Rep2Rep denoising improved image quality and anatomical visualization. The acquisition time was reduced from 3 min 58 s for the clinical protocol without parallel imaging to 2 min 06 s for the high-receiver-bandwidth protocol with 2-fold GRAPPA acceleration.

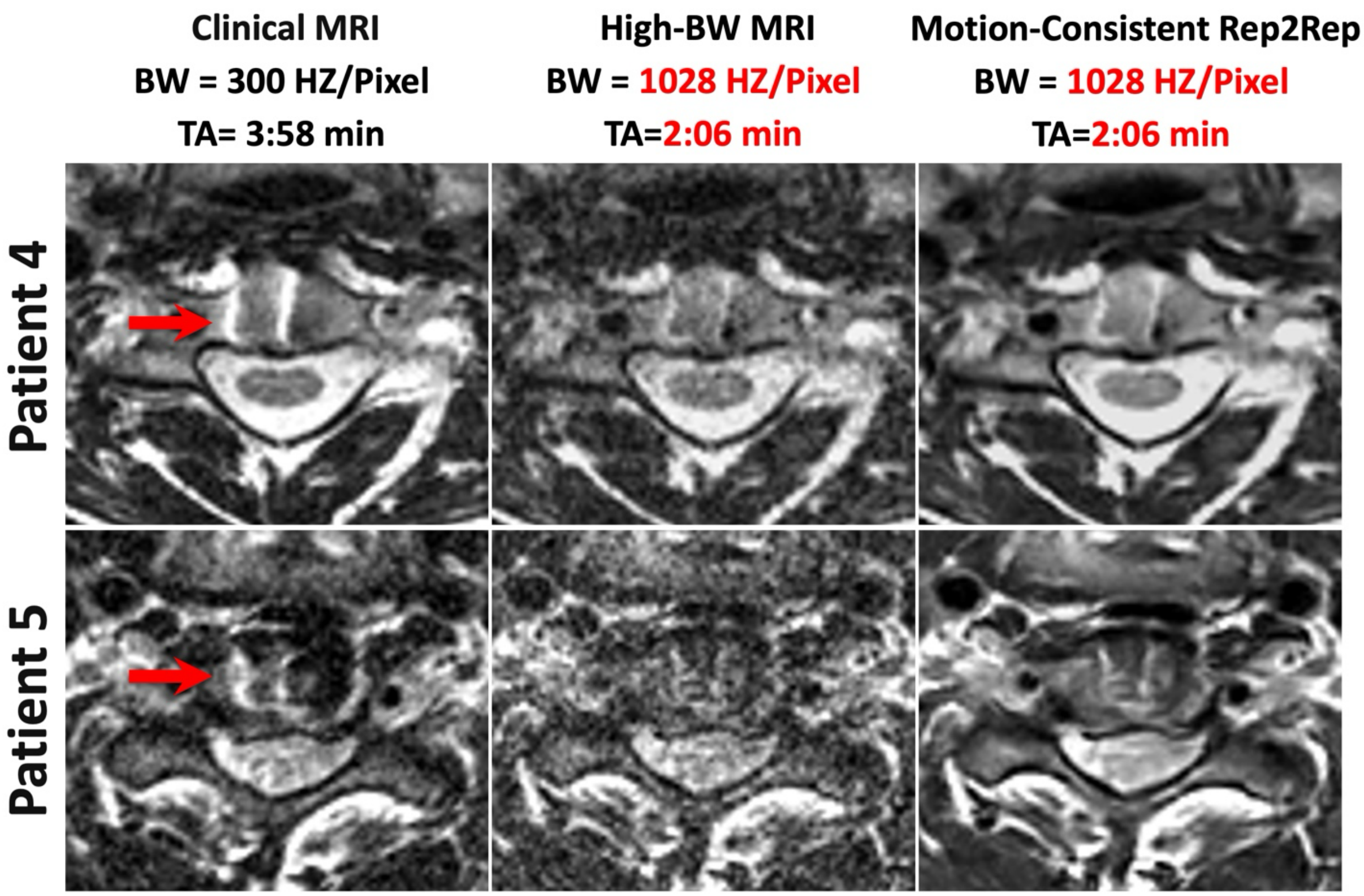

**Figure 5:** Additional examples of high-receiver-bandwidth axial cervical spine MRI from two patients with metal implants. Clinical MRI acquired at 300 Hz/pixel (left column) is compared with high-receiver-bandwidth MRI acquired at 1028 Hz/pixel before (middle column) and after (right column) motion-consistent Rep2Rep denoising. The red arrows indicate metal-induced artifacts on clinical images. High-receiver-bandwidth MRI combined with Rep2Rep denoising yielded reduced metal-induced distortion and improved image quality, as well as reduced acquisition time from 3 min 58 s to 2 min 06 s compared to the clinical protocol.

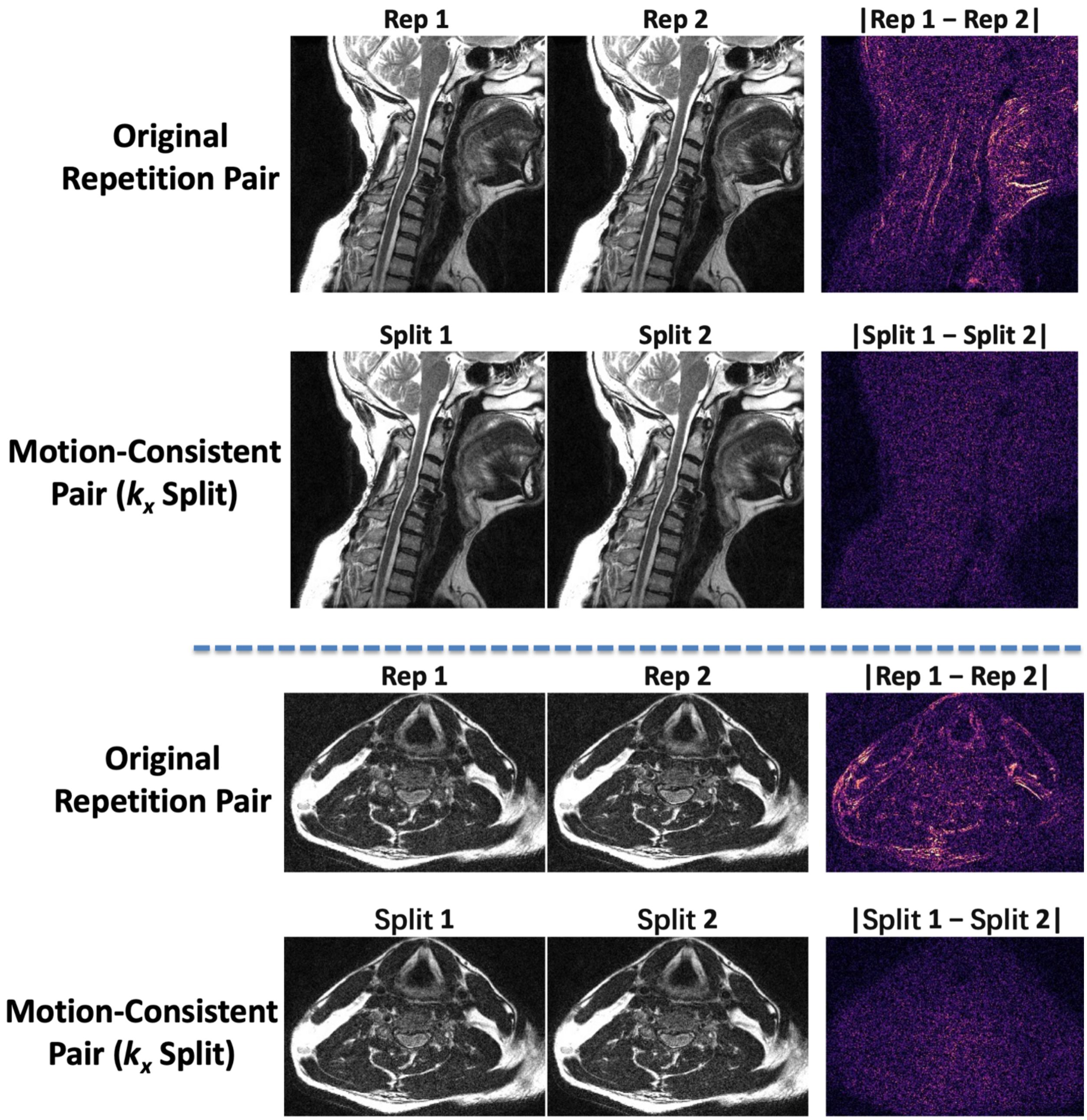

**Figure 6:** Representative sagittal (top) and axial (bottom) cervical spine MRI images comparing the anatomical consistency of the original repetitions and the proposed motion-consistent training repetitions. The original repetitions showed visible anatomical mismatch, while the proposed motion-consistent repetitions showed substantially reduced structural differences, and improved anatomical consistency between the training repetitions.

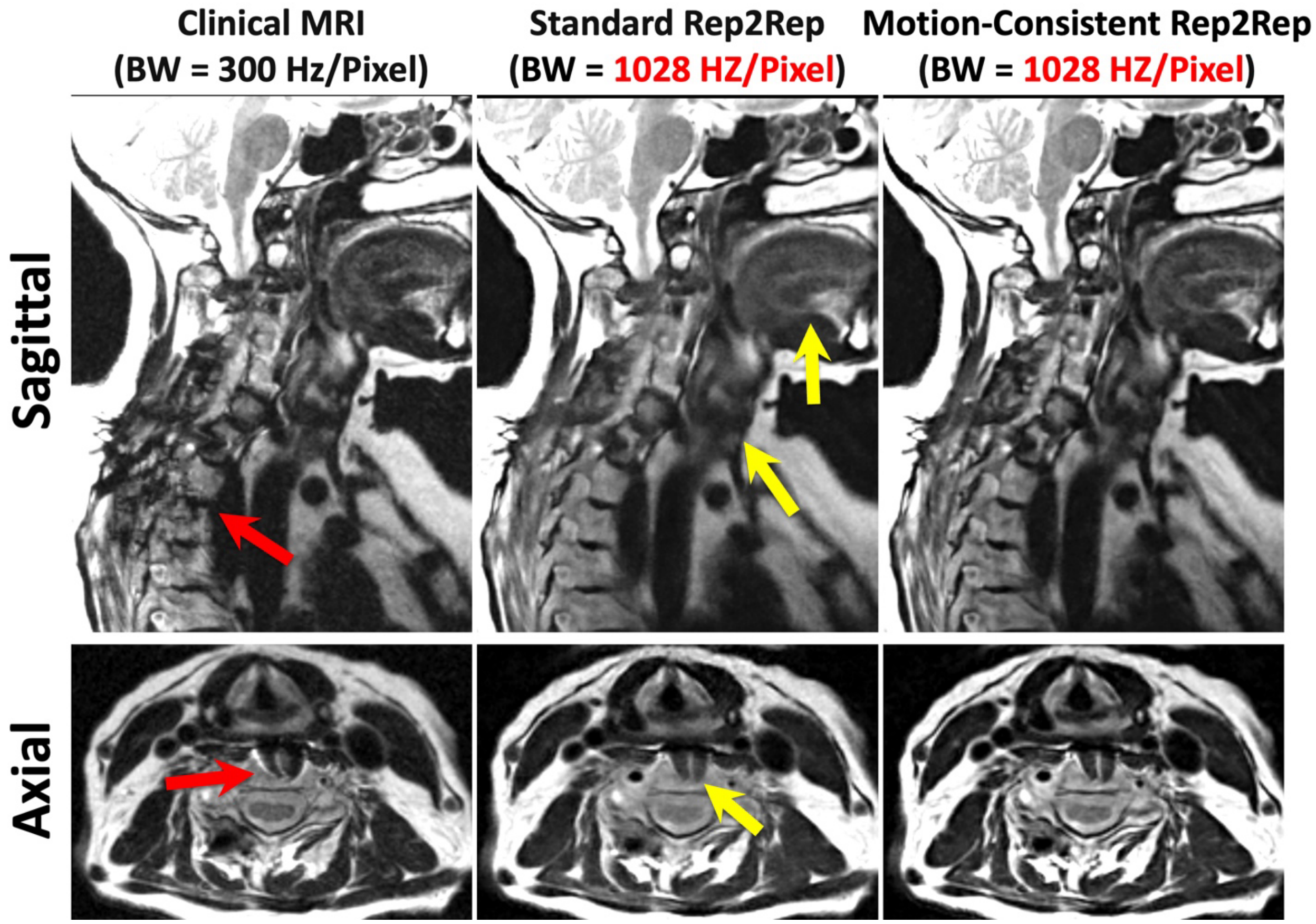

**Figure 7:** Representative sagittal (top row) and axial (bottom row) cervical spine MRI comparing standard and motion-consistent Rep2Rep denoising. Clinical low-receiver-bandwidth images are also included for reference, with red arrows indicating metal-induced artifacts. Compared with standard Rep2Rep, motion-consistent Rep2Rep enabled better preservation of anatomical structures and fine details. Image blurring and loss of fine anatomical details observed in standard Rep2Rep are indicated by the yellow arrows. Both denoising methods underwent the same sharpening and dithering procedures to ensure a fair comparison.

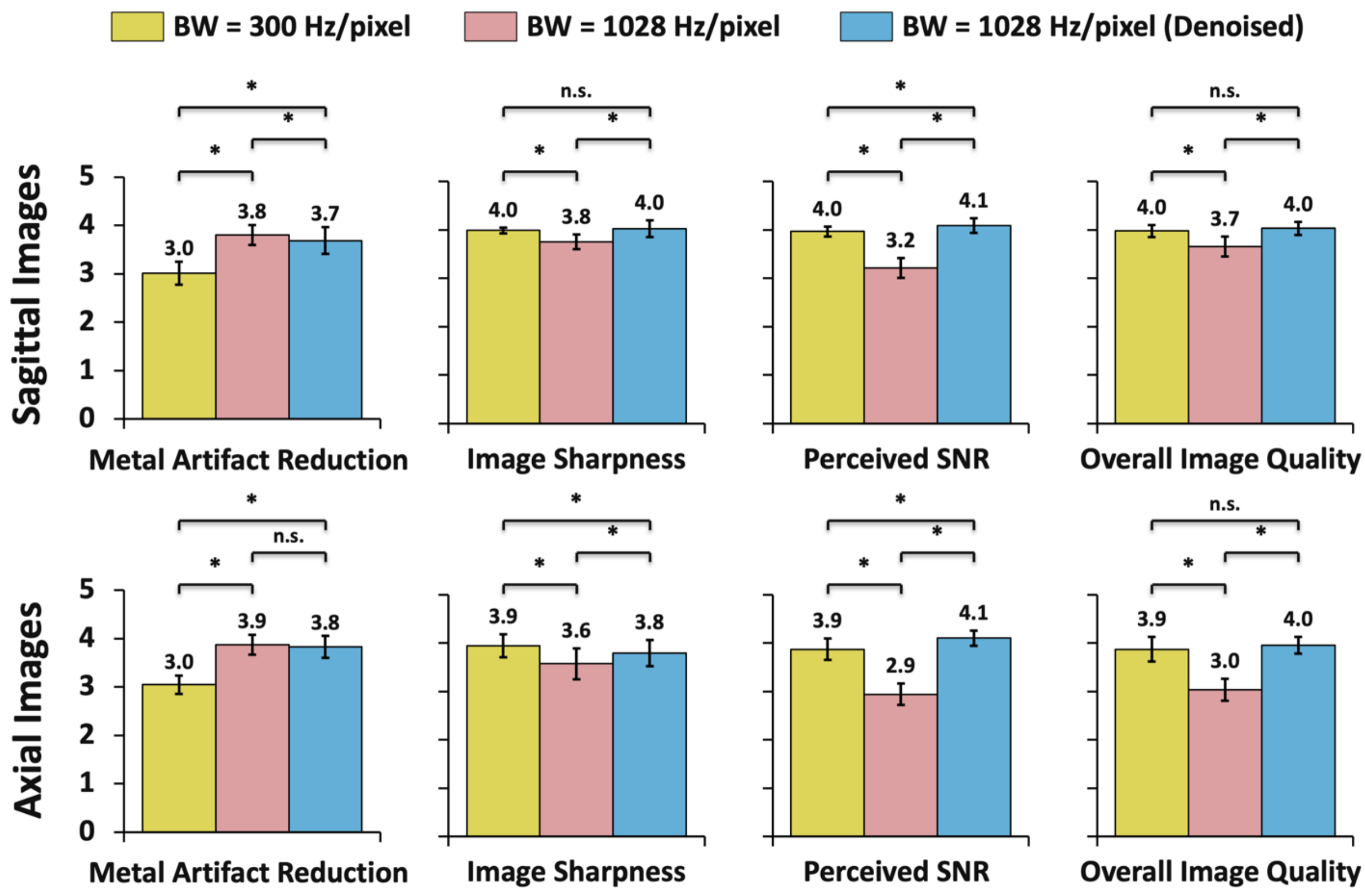

**Figure 8:** Results of the image quality evaluation performed by three radiologists for sagittal (top row) and axial (bottom row) cervical spine MRI. Each metric was assessed using a five-point scale (5-1: best to non-diagnostic). An asterisk indicates a statistically significant difference ($P < 0.05$), while n.s. indicates a non-significant difference.

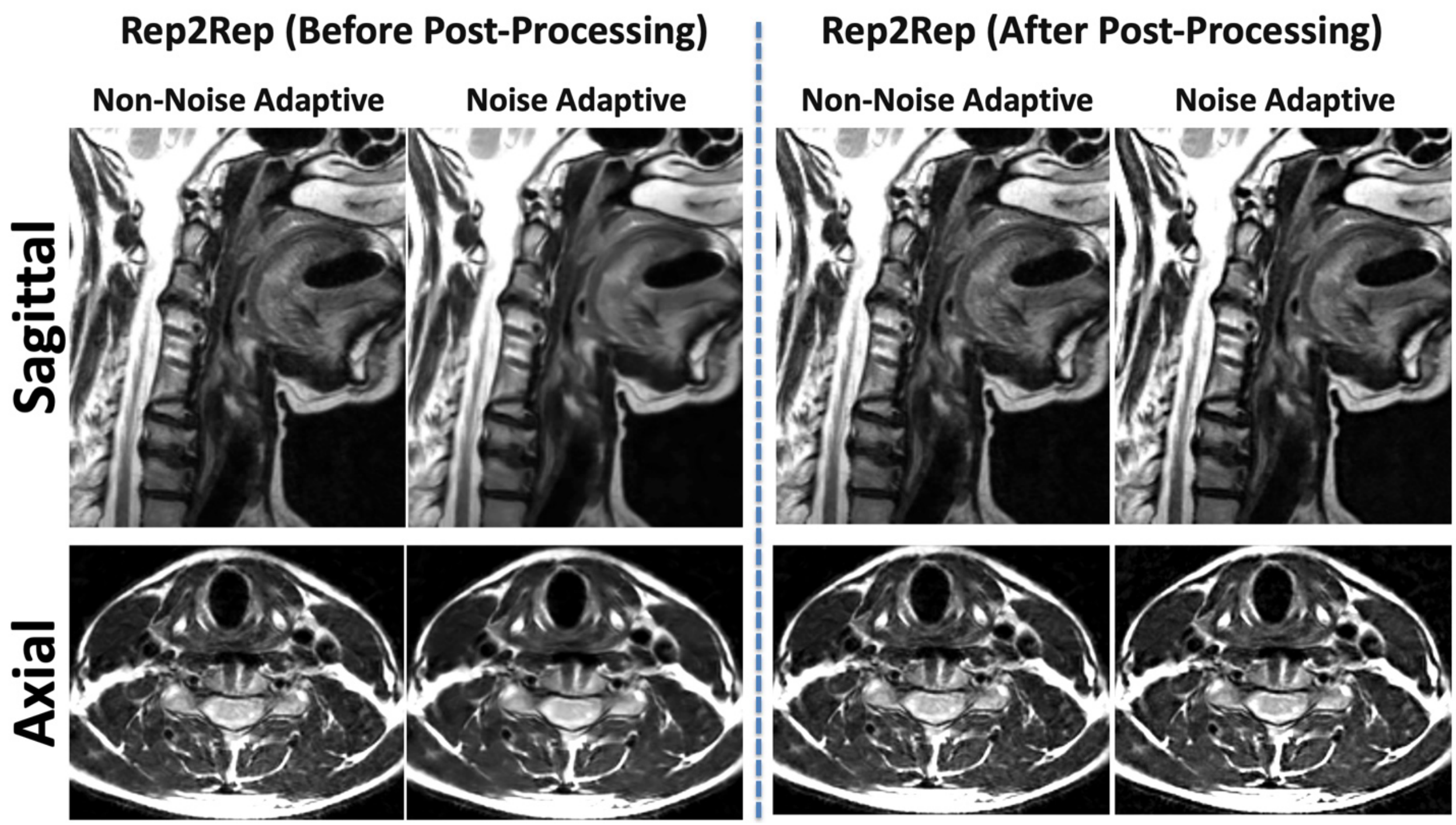

**Figure 9:** Representative sagittal (top row) and axial (bottom row) high-receiver-bandwidth images comparing noise-adaptive and non-adaptive Rep2Rep before (left panel) and after (right panel) post-processing. Before postprocessing, non-adaptive Rep2Rep showed noticeable residual noise. However, after sharpening and dithering, the two approaches yielded visually comparable final images.

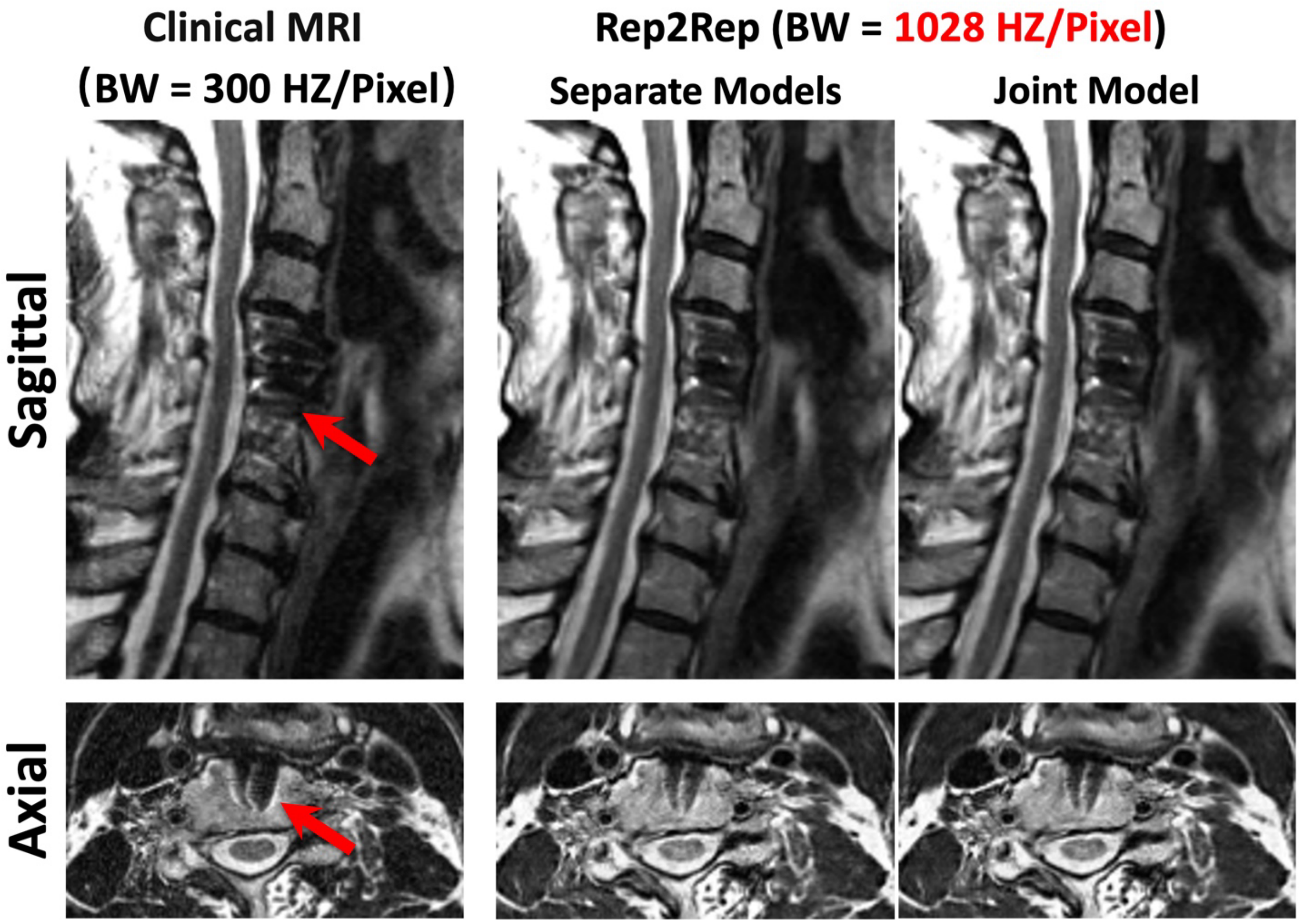

**Figure 10:** Representative sagittal (top row) and axial (bottom row) cervical spine images comparing joint and orientation-specific Rep2Rep training. Both strategies yielded visually comparable results despite differences in image orientations, spatial resolution, and acquisition parameters.

## Supporting Figures

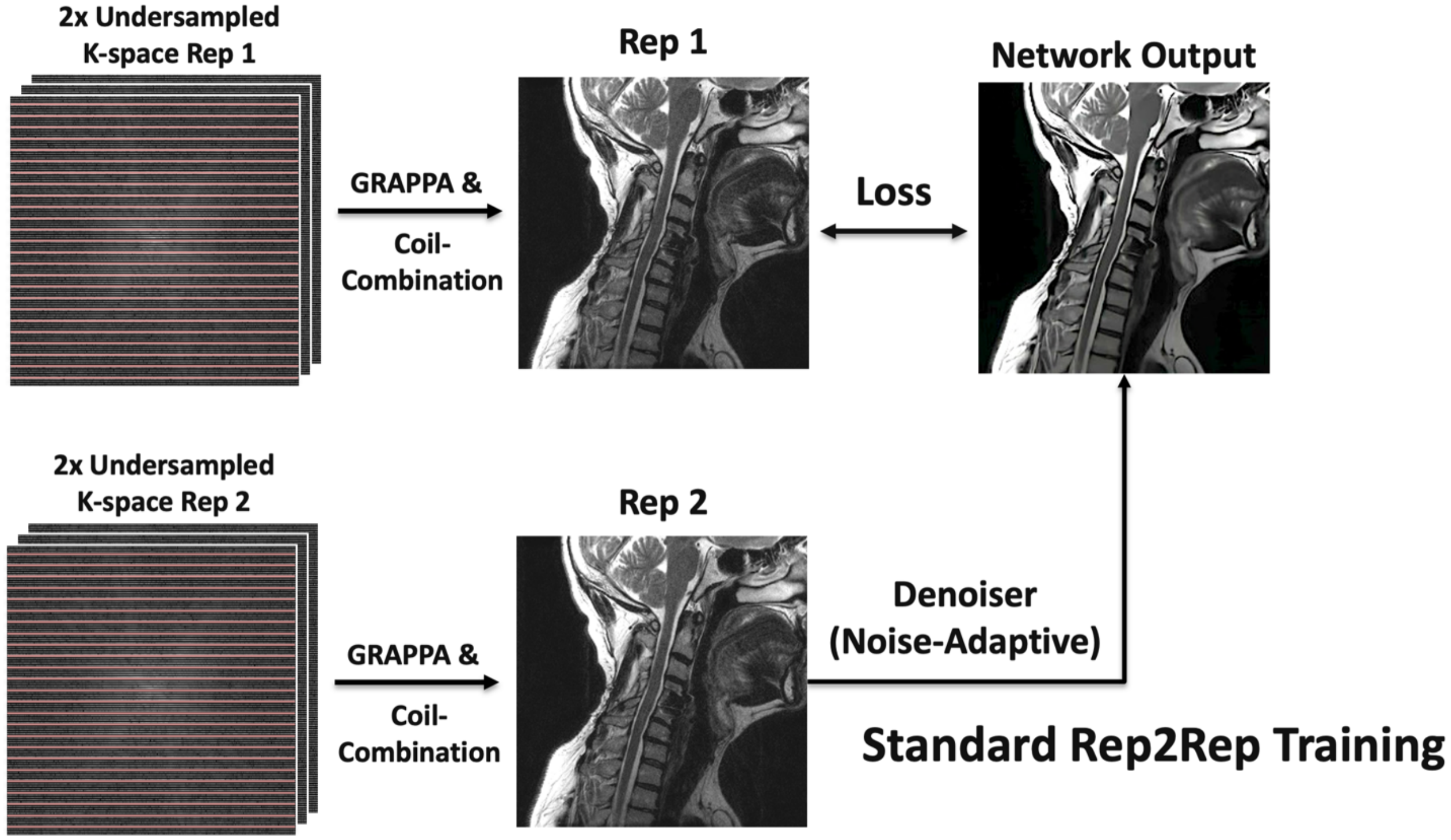

**Supporting Information Figure S1:** In standard Rep2Rep denoising, two independently acquired undersampled k-space repetitions are separately reconstructed using GRAPPA and coil combination to generate the input and target images for self-supervised training. Because the training input and target are reconstructed from independently acquired repetitions, any anatomical mismatch due to inter-repetition motion may affect denoising performance.

| Coil # | Before Readout OS Removal | After Readout OS Removal | Gaussian Noise |
|---|---|---|---|
| 1 | 0.2565 | 0.0959 | 0.0739 |
| 2 | 0.1973 | 0.0547 | 0.0781 |
| 3 | 0.2279 | 0.1172 | 0.0483 |
| 4 | 0.2052 | 0.0925 | 0.1022 |
| 5 | 0.2692 | 0.0199 | 0.0933 |
| 6 | 0.2739 | 0.0539 | 0.0772 |
| 7 | 0.2150 | 0.0286 | 0.0991 |
| 8 | 0.2780 | 0.0962 | 0.0524 |
| 9 | 0.2770 | 0.0099 | 0.0928 |
| 10 | 0.2206 | 0.0293 | 0.0892 |
| 11 | 0.2666 | 0.0380 | 0.0752 |
| 12 | 0.2684 | 0.0346 | 0.0772 |
| 13 | 0.2013 | 0.0965 | 0.0728 |
| 14 | 0.2277 | 0.0797 | 0.0800 |
| 15 | 0.2047 | 0.1086 | 0.0820 |
| 16 | 0.2961 | 0.0511 | 0.0683 |
| **Mean ± SD** | **0.2428 ± 0.0334** | **0.0629 ± 0.0348** | **0.0789 ± 0.0149** |

**Supporting Information Figure S2:** Evaluation of noise correlation between odd and even readout k-space samples before and after removal of readout oversampling. Noise correlation was evaluated using data from a noise-only scan acquired within a sagittal cervical spine MRI dataset (16 receiver coils). The noise data had 512 readout samples per coil, including 2-fold readout oversampling. After removal of readout oversampling, the readout matrix was reduced to 256 samples per coil. For both conditions (with and without oversampling), the k-space data were uniformly divided into two complementary subsets containing odd- and even-indexed samples along the readout direction. The complex correlation coefficient between the two subsets was calculated separately for each receiver coil, and its magnitude was used to quantify noise correlation. The mean correlation was then calculated across all 16 coils. For comparison, independent complex Gaussian noise with the same dimensions as the noise data after oversampling removal (256 readout samples × 16 coils) was generated and evaluated using the same odd/even partition and correlation analysis. The mean ± SD noise correlation was 0.2428 ± 0.0334 before oversampling removal. This correlation is primarily due to vendor-implemented anti-aliasing filters applied along the readout direction, which limit the receiver noise bandwidth and can introduce correlations between adjacent readout samples (28). The mean ± SD noise correlation was reduced to 0.0629 ± 0.0348 after removing the readout oversampling, and this is close to the correlation obtained from independent complex Gaussian noise (0.0789 ± 0.0149). The nonzero correlation magnitude observed for Gaussian noise reflects finite-sample estimation variability. These results demonstrate that removal of readout oversampling substantially reduces noise correlation between the odd and even k-space subsets, and this helps satisfy the approximate noise-independence assumption of the Noise2Noise principle and the proposed motion-consistent Rep2Rep training.

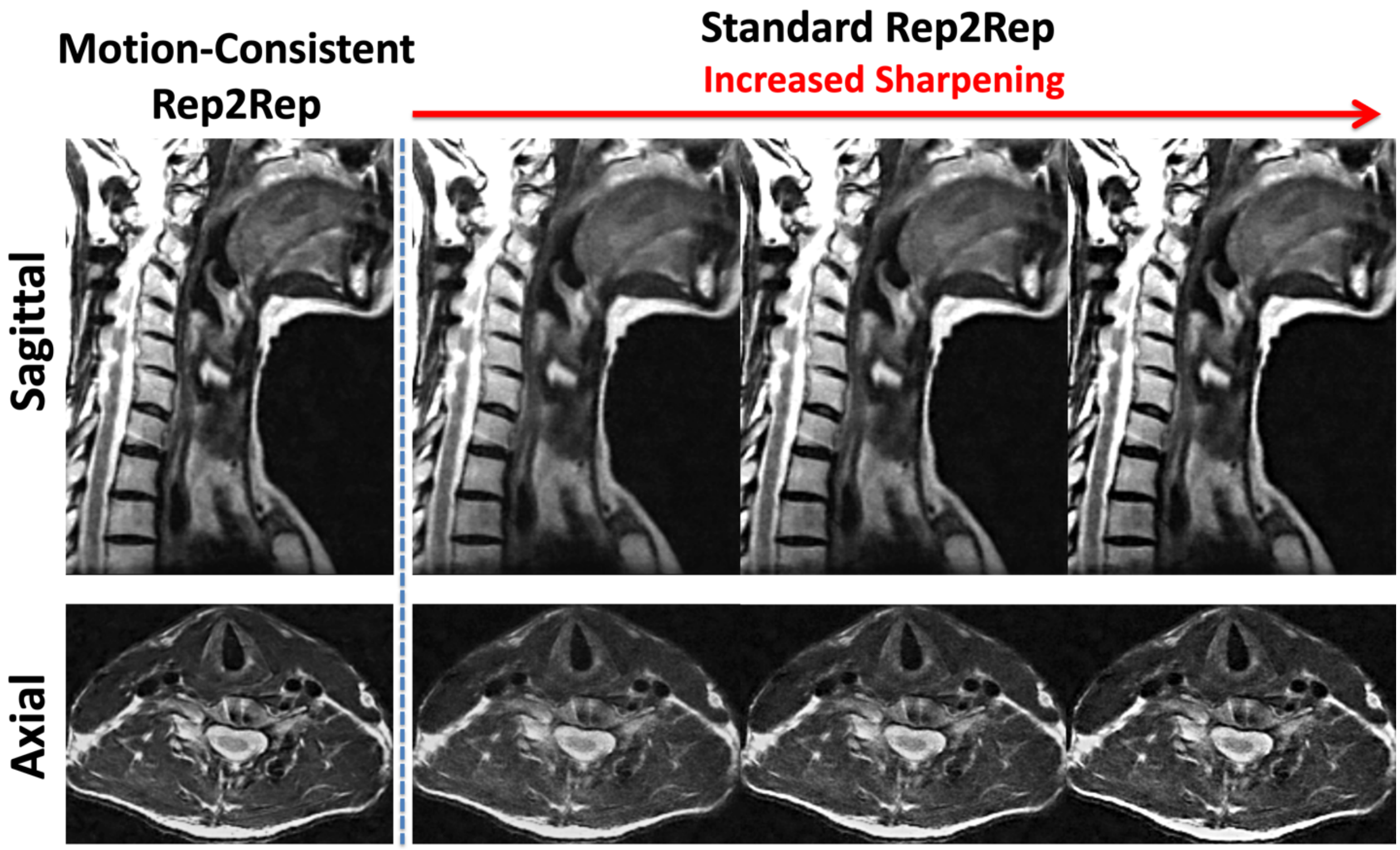

**Supporting Information Figure S3:** Representative sagittal (top row) and axial (bottom row) cervical spine images comparing motion-consistent Rep2Rep with standard Rep2Rep after progressively stronger sharpening. Although increased sharpening improved the apparent sharpness of standard Rep2Rep images, it did not fully restore the fine anatomical details preserved by motion-consistent Rep2Rep.